\newif\ifarxiv\arxivtrue
\newif\ifcdfnote\cdfnotefalse
\newif\ifdisstyle\disstylefalse
\newif\iffnalconf\fnalconffalse

\ifcdfnote
\documentclass[12pt,twoside,letter]{article}
\fi

\ifarxiv
\documentclass[12pt,twoside,letter]{article}
\fi

\iffnalconf
\documentclass[12pt,twoside,letter]{article}
\fi

\ifdisstyle
\documentclass[twoside]{dis08}
\fi

\usepackage[latin1]{inputenc}
\usepackage[dvips]{graphicx,epsfig,color}
\usepackage{wrapfig,rotating}
\usepackage{amssymb,amsmath,array}

\ifdisstyle
\else
\fi

\def\mrm{\mathrm}

\newcommand{\eV}{\ensuremath{\mathrm{e\kern -0.1em V}}}
\newcommand{\TeV}{\ensuremath{\mathrm{Te\kern -0.1em V}}}
\newcommand{\GeV}{\ensuremath{\mathrm{Ge\kern -0.1em V}}}
\newcommand{\MeV}{\ensuremath{\mathrm{Me\kern -0.1em V}}}
\newcommand{\keV}{\ensuremath{\mathrm{ke\kern -0.1em V}}}
\newcommand{\GeVc}{\ensuremath{\GeV/c}}
\newcommand{\GeVcc}{\ensuremath{\GeV/{c^2}}}

\newcommand{\um}{\ensuremath{\mathrm{\mu m}}}
\newcommand{\fm}{\ensuremath{\mathrm{fm}}}
\newcommand{\mm}{\ensuremath{\mathrm{mm}}}

\newcommand{\pbin}{\ensuremath{\mathrm{pb}^{-1}}}
\newcommand{\fbin}{\ensuremath{\mathrm{fb}^{-1}}}

\newcommand{\ppbar}{\ensuremath{p\overline{p}}}

\newcommand{\qqbar}{\ensuremath{q\overline{q}}}

\newcommand{\tevE}{\ensuremath{\sqrt{s} = 1.96~\TeV}}
\newcommand{\Mpl}{\ensuremath{M_{\mrm{Pl}}}}
\newcommand{\MD}{\ensuremath{M_D}}

\newcommand{\etaphi}{\text{$\eta$-$\phi$}}

\newcommand{\met} {\mbox{${E\!\!\!\!/_T}$}}

\newcommand{\gmet}{\ensuremath{\gamma+\met}}
\newcommand{\jmet}{\ensuremath{\mrm{jet}+\met}}

\begin{document}

\title{
\ifcdfnote
\begin{flushright}\large{
CDF/PHYS/EXOTIC/PUBLIC/9377\\
\today
}
\end{flushright}
\vspace{1cm}
\fi
\iffnalconf
\begin{flushright}\large{
FERMILAB-CONF-08-218-E\\
\today
}
\end{flushright}
\vspace{0.25cm}
\fi
Searches for Large Extra Dimensions at the Tevatron
}

\author{Vyacheslav Krutelyov$^1$ for the CDF and the D\O\ Collaborations
\iffnalconf
\thanks{Work supported by the U.S. Department of Energy under contract No. DE-AC02-07CH11359.}
\fi
\ifarxiv
\thanks{Work supported by the U.S. Department of Energy under contract No. DE-AC02-07CH11359.}
\fi
%
%
\vspace{.3cm}\\
%
1- University of California at Santa Barbara - Dept of Physics \\
Santa Barbara CA 93106-9530 - U.S.A.\\
%
}
\ifarxiv
\date{\vspace{0.5cm}}
\fi

\maketitle

\begin{abstract}
The presence of extra dimensions can be probed in high energy collisions
via the production or exchange of gravitons.
The former corresponds to signatures with missing energy while the latter
corresponds to modifications of the final state spectra.
Here I review results of analyses performed by the CDF and D\O\ Collaborations
on \ppbar\ collisions at \tevE\ in signatures sensitive 
to large extra dimensions.
These include analyses of \gmet\ and \jmet\ as signatures of graviton production
as well as analyses of dilepton and diboson final states sensitive
to graviton exchange.
\end{abstract}

\section{Introduction}
The standard model of particle physics (SM) does not account for gravitational interactions.
A number of theoretical models beyond the SM
naturally include gravity and require the existence
of extra dimensions where only gravity propagates freely.
In particular, a model proposed in Ref.~\cite{ADD}
assumes there are large extra dimensions (LED) compactified on a sub-millimeter scale $R$.
In this model gravity can become strong at the \TeV\ scale,
which effectively solves the hierarchy problem present in the SM.
For $n$ extra dimensions
the four-dimensional Planck mass \Mpl\ is related to the $4+n$-dimensional
fundamental scale \MD\ by $\Mpl^2 = 8\pi \MD^2 (\MD R)^n$.
For \MD\ in the \TeV\ range  $R$ is of the order of 0.1~\mm\ (10~\fm) for $n=2$ ($6$).
To solve the hierarchy problem \MD\ should not be too large, which rules out  $n=1$.
Tests of the Newton's law constrain $R$ to be below $37~\um$ for $n=2$~\cite{PDG}.
Other constraints come from astrophysics (up to $n=3$)
and from high energy colliders~\cite{PDG}.

The cross sections of the LED processes depend on $M_D$ and can be at
levels detectable at the Tevatron~\cite{GRW}.
In \ppbar\ collisions the  graviton can be produced in
$\qqbar\to g G$, $qg \to q G$, or $gg\to g G$ corresponding to a \jmet\ 
final state, and  
$\qqbar \to \gamma G$ corresponding to a \gmet\ detector signature~\cite{unitDefs}.
Graviton exchange can be studied in a range of
$2\to 2$ processes with the best sensitivity  in final states with
two leptons, photons, or $Z$ bosons.

In this review~\cite{DISslides} I first describe searches for graviton production
in the \gmet\ and \jmet\ signatures reported by CDF and
in the \gmet\ signature reported by D\O.
Then I briefly describe searches sensitive to graviton exchange
in dimuon, dielectron and diphoton final states reported by D\O\ and
in $ZZ$ final states by both CDF and D\O.


\section{CDF and D\O\ detectors}
\label{sec:detectors}
A detailed description of the CDF and the D\O\ detectors can be found in Ref.~\cite{tdrCDF,tdrD0}.
The vertexing and the tracking detectors surrounding the interaction region
are used to reconstruct charged particle trajectories.
Further out are the calorimeters with electromagnetic and hadronic
longitudinal segmentation
used to identify and measure  photons, electrons, and jets.
Outside the calorimeter
are the muon detectors used to identify muons from the collisions and cosmic rays.

In analyses of \gmet\ events one of the major backgrounds is from cosmic ray
muons misidentified as photons.
The detector features used to reject this background are:
{\em photon pointing} available at D\O\ allowing the reconstruction of direction
of the photon from the longitudinal segmentation of the electromagnetic calorimeter~\cite{D0IIgmet};
{\em photon timing} available at CDF from the EMTiming system~\cite{EMTnim}
providing the measurement of the time of the energy deposit 
and allowing to suppress cosmic-ray background by a factor of 20 or more.


\section{Searches for graviton production}
The production of gravitons in hadron collisions
is probed in events where the final-state hadronic jet or photon recoils against
the graviton, which is not detected and results in \met.
\begin{wraptable}{r}{0.36\columnwidth}
\centerline{\begin{tabular}{|l|c|}
\hline
Background	& Events 	\\ \hline
$Z\to\nu\nu$	& $ 388 \pm 30$ \\
$W\to \ell\nu$	& $ 362 \pm 17$	\\
Multi-jet	& $ 23 \pm 20$	\\
$\gamma+$jet	& $ 17 \pm 5$	\\
Non-collision	& $ 10 \pm 10$	\\
Total predicted & $ 808 \pm 62$ \\
Data observed 	& $ 809 $	\\
\hline
\end{tabular}}
\caption{The numbers of expected background and  observed events
in the \jmet\ sample used by CDF.}
\label{tab:jmetEvts}
\end{wraptable}
Searches in both the \jmet\ and the \gmet\ require exclusive signatures:
only a jet  or a photon is in the event with a veto on the presence of other
objects.
These searches have similar backgrounds.
The dominant and only irreducible background is production of $Z\to\nu\nu$ in 
association with a photon or a jet.

A search for LED in \jmet\ events has been performed by CDF using 1.1~\fbin\ of data,
updating the analysis in Ref.~\cite{CDFjmet360}.
Events with high \met, a high-$E_T$ jet and no second jet with $E_T > 60~\GeV$
are selected.
All major backgrounds are estimated in a data-driven way.
The contributions from $Z\to\nu\nu$ and from $W\to\ell\nu$, where the lepton
is lost, are estimated using measured $Z\to\ell\ell$ and $W\to\ell\nu$ events.
After an {\em a priori} optimization for the best sensitivity to LED,
events with the leading jet $E_T>150~\GeV$ and $\met > 120~\GeV$ are selected.
The summary of predicted and observed events is given in Table~\ref{tab:jmetEvts}.
The constraints on LED are given at the end of this section.

\begin{wraptable}{l}{0.5\columnwidth}
\centerline{\begin{tabular}{|l|c|c|}
\hline
Background	& CDF 			& D\O 		\\ \hline
$Z\gamma\to\nu\nu\gamma$	& $ 24.8 \pm 2.8$ 	& $12.1 \pm 1.3$\\
$W\to (e\to\gamma)\nu$& $2.6 \pm 0.4$	& $3.8 \pm 0.3$	\\
$W\to(\mu/\tau\to\gamma)\nu$& $1.0\pm 0.2$	& --		\\
$W\gamma\to \ell \nu\gamma$	& $ 5.0 \pm 1.4$	& $1.5 \pm 0.2$	\\
jet$\to\gamma$	& --			& $ 2.2 \pm 1.5$\\
$\gamma\gamma\to\gamma$	& $ 2.3 \pm 0.6$& -	\\
Non-collision	& $ 9.8 \pm 1.3$	& $2.8 \pm 1.4$	\\
Total predicted & $ 46.3 \pm 3.0$ 	& $22.4 \pm 2.5$ \\
Observed data 	& $ 40 $		& $29$	\\
\hline
\end{tabular}}
\caption{The numbers of expected background and observed events in the \gmet\ samples 
analyzed by CDF and D\O. The contribution from jet$\to\gamma$
in CDF is included in $(W/\gamma)\gamma$  by virtue of the background
estimation method.}
\label{tab:gmetEvts}
\end{wraptable}
Results for the \gmet\ signature in the Tevatron Run~II 
have been recently reported by CDF in 2~\fbin\ of data and by D\O\ in 1.05~\fbin\ of 
data~\cite{D0IIgmet}.
Both analyses use events with high \met\ and a high-$E_T$ photon in 
the central detector region with $|\eta|<1.1$~\cite{unitDefs}.
No jets with $E_T>15~\GeV$ are allowed in the event in order to suppress QCD backgrounds.
Also, no high-$p_T$ track is allowed, suppressing contributions  from leptonic $W$
and $Z$ decays: tracks with $p_T>10~\GeVc$ (isolated with $p_T>6.5~\GeVc$) are vetoed
by CDF (D\O).
The contribution from cosmic rays is suppressed based on presence of hits in the muon detectors
as well as using the {\em photon pointing} or the {\em photon timing}
mentioned in Section~\ref{sec:detectors}.

The background contributions for both analyses are summarized in Table~\ref{tab:gmetEvts}.
The number of events from $Z\to\nu\nu$ is estimated from simulation.
The contribution from $W\to e\nu$, where the electron is identified as 
the photon,
is extracted from the number of events with electrons
passing all other requirements applied to the photon signal sample, scaled
by the misidentification rate.
Contributions from cosmic rays and jets misidentified as photons are estimated at D\O\ using
the {\em photon pointing} method.
The remaining backgrounds in the analysis by D\O\ are estimated using simulation.
The contribution from cosmics at CDF is estimated using events with the photon time
 significantly different from the collision time.
The remaining backgrounds at CDF are  from processes where an object is lost.
Except for $W\to\tau\nu$, which is estimated from simulation,
each of these backgrounds is given by the  number of events in data with
this object identified, scaled by the simulated rate for it to be lost.
This approach implicitly includes contributions from a jet identified as a photon.

\begin{table}[hbt]
\centerline{\begin{tabular}{|l|c|c|c|c|c|}
\hline
Source		& LEP	& D\O	& \multicolumn{3}{c|}{CDF} \\
$n$		& \gmet	& \gmet	& \jmet	& \gmet	& combined		\\ \hline
2	 	& 1600	& 921	& 1310	& 1080	& 1400	 	\\
3	 	& 1200	& 877	& 1080	& 1000	& 1150	 	\\
4	 	& 940	& 848	& 980	& 970	& 1040	 	\\
5	 	& 770	& 821	& 910	& 930	& 980	 	\\
6	 	& 660	& 810	& 880	& 900	& 940	 	\\
\hline
\end{tabular}}
\caption{Lower limits on $M_D$ in \GeVcc\ at $95\%$ C.L. for $n$ from 2 to 6
observed in the \jmet\ and \gmet\ signatures and the combination 
of the two at CDF,
 and in the \gmet\ signature at D\O, along
with the constraints from the LEP experiments~\cite{LEDlep}.
Constraints provided by D\O\ for $n$ above 6 can be found in~\cite{D0IIgmet}.
}
\label{tab:gjmetLimits}
\end{table}
After an {\em a priori} optimization for the best sensitivity,
events with a  photon with $E_T>90~\GeV$ and $\met>50 (70)~\GeV$ are selected
in the CDF (D\O) signal sample.
Event counts observed in the data are consistent with expectations from the backgrounds.
The constraints on the LED model are summarized in Table~\ref{tab:gjmetLimits}.
Since the sensitivity in the \gmet\ is similar to that in the \jmet\ mode,
the constraints can be improved in a combination.
The result of this combination using the CDF data is shown in Table~\ref{tab:gjmetLimits}.
Depending on the number of extra dimensions, the sensitivity to LED
at the Tevatron is comparable to or better than that from the LEP experiments~\cite{LEDlep}.

\section{Searches for virtual graviton exchange}
In hadron collisions the processes most sensitive to  virtual graviton exchange
in LED are those with two leptons (electrons or muons), two photons,
or two $Z$ bosons in the final state, where it is
possible to reconstruct the full kinematics of the final state with high precision.
The value of the graviton exchange amplitude 
depends on the cutoff energy $\Lambda$, presumably of order \MD~\cite{GRW}.
The cross section is proportional to $\Lambda^{-8}$.

A search for LED in the dimuon final state has been reported by D\O\ using 200~\pbin\ of
data~\cite{LEDmumuD0}.
The sample is comprised of events with two opposite charge muons with 
$p_T$ above $15~\GeVc$ and both muons passing additional quality and isolation
requirements.
The dilepton invariant mass and the scattering angle in the dilepton
center of mass are analyzed simultaneously to improve sensitivity 
to the exchange of a graviton (spin 2 particle).
A similar analysis using a {\em di-em} final state where no discrimination
is made between electrons and photons has been performed at D\O\ with 240~\pbin\ of
 data~\cite{LEDggD0}.
In this case events were selected if they had two {\em em} objects with $E_T>25~\GeV$ 
passing quality and isolation requirements.
No excess over the expected backgrounds is observed.
The constraints on the parameter $\Lambda$ are in the range of 1 to 2~\TeV\ depending
on the theoretical parameters.

Both CDF and D\O\ have recently reported on searches for $\ppbar\to ZZ$ production
in a four-lepton final state~\cite{CDFzz,D0zz}.
The observed number of events is consistent with the SM expectations.
Although no constraint has been provided on LED by either experiment, based on
Ref.~\cite{LEDzzTh} the constraint on $\Lambda$ is expected to  be in the range of 1.5 to 2.5~\TeV.

\section{Summary}
Searches for LED in signatures with direct graviton production
and those with virtual graviton exchange are explored by the CDF and the D\O\ experiments.
No evidence of LED has been seen.
Additional \ppbar\ collision data currently produced at the Tevatron
is expected to give more insight about the possible presence of large extra spacial dimensions.





\section*{Acknowledgements}
I would like to acknowledge the funding institutions supporting
the CDF and D\O\ Collaborations.
The full list of agencies can be found in, e.g,~\cite{tdrCDF,tdrD0}.

\begin{footnotesize}

\end{footnotesize}



\begin{thebibliography}{99}

\bibitem{ADD}
  N.~Arkani-Hamed, S.~Dimopoulos and G.~R.~Dvali,
  Phys.\ Lett.\  B {\bf 429}, 263 (1998)
  [arXiv:hep-ph/9803315].

\bibitem{PDG}
  W.~M.~Yao {\it et al.}  [Particle Data Group],
  J.\ Phys.\ G {\bf 33}, 1 (2006).

\bibitem{GRW}
  G.~F.~Giudice, R.~Rattazzi and J.~D.~Wells,
  Nucl.\ Phys.\  B {\bf 544}, 3 (1999)
  [arXiv:hep-ph/9811291].


\bibitem{unitDefs}
The polar coordinate system is used with an origin at the center of a detector
and the $z$-axis ($\theta=0$) along the proton beam.
The pseudorapidity $\eta$ is defined as $\eta=-\ln[\tan(\theta/2)]$.
The transverse momentum $p_T$ is defined as $p_T=p\cdot\sin{\theta}$,
where $p$ is the particle's momentum. 
The transverse energy is defined as $E_T = E\cdot\sin{\theta}$, 
where $E$ is the energy measured by the calorimeter.
The missing $E_T$ (\met) is defined as $\vec{\met} = - \sum_i E_T^i \hat{n}_i$,
where $i$ is the index of a calorimeter tower (a segmentation unit in \etaphi) 
and $\hat{n}_i$ is a unit vector perpendicular to the beam axis
pointing to the tower from the origin.

\bibitem{DISslides} Slides: \\
\verb$http://indico.cern.ch/contributionDisplay.py?contribId=104&sessionId=15&confId=24657$


\bibitem{tdrCDF}
  A.~Abulencia {\it et al.}  [CDF Collaboration],
  J.\ Phys.\ G {\bf 34}, 2457 (2007)
  [arXiv:hep-ex/0508029].



\bibitem{tdrD0}
  V.~M.~Abazov {\it et al.}  [D0 Collaboration],
  Nucl.\ Instrum.\ Meth.\  A {\bf 565}, 463 (2006)
  [arXiv:physics/0507191].


\bibitem{D0IIgmet}
  V.~M.~Abazov {\it et al.}  [D0 Collaboration],
  arXiv:0803.2137 [hep-ex], accepted by Phys.\ Rev.\ Lett.

\bibitem{EMTnim}
  M.~Goncharov {\it et al.},
  Nucl.\ Instrum.\ Meth.\  A {\bf 565}, 543 (2006)
  [arXiv:physics/0512171].


\bibitem{CDFjmet360}
  A.~Abulencia {\it et al.}  [CDF Collaboration],
  Phys.\ Rev.\ Lett.\  {\bf 97}, 171802 (2006)
  [arXiv:hep-ex/0605101].

\bibitem{LEDlep}
  J.~Abdallah {\it et al.}  [DELPHI Collaboration],
  Eur.\ Phys.\ J.\  C {\bf 38}, 395 (2005)
  [arXiv:hep-ex/0406019];
  P.~Achard {\it et al.}  [L3 Collaboration],
  Phys.\ Lett.\  B {\bf 587}, 16 (2004)
  [arXiv:hep-ex/0402002].

\bibitem{LEDmumuD0}
  V.~M.~Abazov {\it et al.}  [D0 Collaboration],
  Phys.\ Rev.\ Lett.\  {\bf 95}, 161602 (2005)
  [arXiv:hep-ex/0506063].

\bibitem{LEDggD0}
  V.~M.~Abazov {\it et al.}  [D0 Collaboration],
  D\O\ note 4336 (2004), unpublished.

\bibitem{CDFzz}
  T.~Aaltonen {\it et al.}  [CDF Collaboration],
  arXiv:0801.4806 [hep-ex].

\bibitem{D0zz}
  V.~M.~Abazov {\it et al.}  [D0 Collaboration],
  Phys.\ Rev.\ Lett.\  {\bf 100}, 131801 (2008)
  [arXiv:0712.0599 [hep-ex]].

\bibitem{LEDzzTh}
  M.~Kober, B.~Koch and M.~Bleicher,
  Phys.\ Rev.\  D {\bf 76}, 125001 (2007)
  [arXiv:0708.2368 [hep-ph]] and references therein.


\end{thebibliography}
\end{document}